\documentclass[aps,prd,twocolumn,nofootinbib,floatfix,preprintnumbers]{revtex4}
\usepackage{subfigure}
\usepackage{graphicx}
\usepackage{dcolumn}
\usepackage{epsfig}
\usepackage{amsmath,bm}
\usepackage{amsfonts}
\usepackage{amssymb}
\usepackage{color}
\usepackage{hyperref}
\usepackage[dvipsnames]{xcolor}
\usepackage{orcidlink}

\newcommand{\be}{\begin{equation}}
\newcommand{\ee}{\end{equation}}
\newcommand{\bea}{\begin{eqnarray}}
\newcommand{\eea}{\end{eqnarray}}

\begin{document}

\title{Eigen-decomposition of Covariance matrices : An application to the BAO Linear Point}

\author{Jaemyoung (Jason) Lee,\orcidlink{0000-0001-6633-9793}}
\email{astjason@sas.upenn.edu}
\affiliation{Department of Physics and Astronomy, University of Pennsylvania, Philadelphia, PA 19104, U.S.A. }

\author{Farnik Nikakhtar,\orcidlink{0000-0002-3641-4366}}
\affiliation{Department of Physics, Yale University, New Haven, CT 06511, U.S.A } 

\author{Aseem Paranjape,\orcidlink{0000-0001-6832-9273}}
\affiliation{Inter-University Centre for Astronomy and Astrophysics, Ganeshkhind, Pune 411007, India} 

\author{Ravi K.~Sheth,\orcidlink{0000-0002-2330-0917}}
\affiliation{Center for Particle Cosmology, University of Pennsylvania, Philadelphia, PA 19104 -- USA}

\date{\today}

\begin{abstract}
The Baryon Acoustic Oscillation (BAO) feature in the two-point correlation function (TPCF) of discrete tracers such as galaxies is an accurate standard ruler.~The covariance matrix of the TPCF plays an important role in determining how the precision of this ruler depends on the number density and clustering strength of the tracers, as well as the survey volume.~An eigen-decomposition of this matrix provides an objective way to separate the contributions of cosmic variance from those of shot-noise to the statistical uncertainties.~For the signal-to-noise levels that are expected in ongoing and next-generation surveys, the cosmic variance eigen-modes dominate.~These modes are smooth functions of scale, meaning that:~they are insensitive to the modest changes in binning that are allowed if one wishes to resolve the BAO feature in the TPCF;~they provide a good description of the correlated residuals which result from fitting smooth functional forms to the measured TPCF;~they motivate a simple but accurate approximation for the  uncertainty on the Linear Point (LP) estimate of the BAO distance scale.~This approximation allows one to quantify the precision of the BAO distance scale estimate without having to generate a large ensemble of mock catalogs and explains why:~the uncertainty on the LP does not depend on the functional form fitted to the TPCF or the binning used; 
the LP is more constraining than the peak or dip scales in the TPCF; 
the evolved TPCF is less constraining than the initial one, so that reconstruction schemes can yield significant gains in precision.  
\end{abstract}



\maketitle

\newcommand{\ste}[1]{\textcolor{red}{\textbf{\small[Ste: #1]}}}


\section{Introduction}\label{sec:intro}
Some of the tightest constraints on the cosmological distance - redshift relation come from the BAO feature in the pair correlation function \cite{eisenstein2005bao, anselmi2018CF_standard_ruler, eBOSS2021, abbott2024dark, DESI2024VI_Cosmology}.~This has led to interest in the precision with which the pair correlation function (TPCF) can be measured, and how this precision translates into uncertainties on the distance scale estimate.~As a result, there is significant interest in understanding the covariance between pair counts on different scales.  

On the $\sim 140$~Mpc scales of most interest to BAO cosmology, the Gauss-Poisson approximation to the covariance is rather accurate \cite{smith2008_eppur, grieb2016gaussian_covariance}.~In this approximation, three different terms contribute:~one is a purely cosmological term, the other is a pure shot-noise term, and the third is a combination of the two.~The first part of our paper is devoted to a study of the relative importance of these terms.~We address this by rotating the covariance matrix into diagonal form, checking how many eigenvectors contribute significantly to the total covariance, and then looking at those eigenvectors.~This provides a simple way of seeing which term dominates and when, as well as for understanding the shapes of the eigenvectors.~We use this insight to explore how binning of the pair counts (width of the rectangular bins, or different bin shapes) affects the structure of the covariance matrix.~Our methodology is similar in spirit to the Karhunen-Lo{\`e}ve decomposition \cite{karhunen1947}, in which a stochastic process is represented by orthonormal basis functions and uncorrelated random coefficients. The Karhunen-Lo{\`e}ve decomposition was applied to Cosmic Microwave Background (CMB) maps \cite{bond1995cobe_eigenmode,Bunn_Sugiyama1995,bunn1996calculation} as well as galaxy redshift surveys \cite{Vogeley1996_eigen_galaxy} in the 1990s for optimal data compression around the time of the first large-scale cosmological surveys.~\cite{tegmark1997karhunen} provide a generalization of this method in anticipation of subsequent generation surveys where the amount of data collected is too large for an uncompressed maximum likelihood analysis.~They found that for both CMB maps and redshift-space distortions, a compression by a factor of $\sim 10$ is achievable by keeping the first $\sim 10\%$ of eigen-modes from the covariance matrix.

The second half of this paper applies these insights to a particular estimator of the BAO scale: the Linear Point (LP).~The LP feature in the correlation function of dark matter or galaxies can be used as a standard cosmological ruler \cite{anselmi2016LP_BAO, anselmi2019LP_fitting, anselmi2018CF_standard_ruler, parimbelli2021LP_neutrino, O'Dwyer2020LP_standard_ruler, nikakhtar2021BAO_Laguerre, BAO_LP_MG,paranjape2022,he2023_BOSS_LP}. The LP lies midway between the peak and dip values in $\xi(r)$, the two-point correlation function:
\begin{equation}
 r_{\rm LP}\equiv \frac{r_{\rm peak} + r_{\rm dip}}{2}.  
 \label{eq:rLP}
\end{equation}
Evidently, the precision with which $r_{\rm LP}$ can be estimated from data depends on the covariance between the $r_{\rm peak}$ and $r_{\rm dip}$ estimates.~In turn, this depends on the covariance matrix of the measurements, which depends on the widths of bins in which pairs were counted (or, more generally, the bin shapes themselves).~This has led to significant computational efforts simply to determine the optimal bin width \cite{systematics2014, anselmi2018LP_mock_catalogs}.

Moreover, $r_{\rm LP}$ is typically estimated by fitting a pre-determined functional form to the measured $\xi$ (e.g. polynomials, Chebyshev polynomials, generalized half-integer Laguerre functions).~The associated error bars would then appear to be closely tied to this functional form (e.g.~Equations~2.6 and~9 in \cite{parimbelli2021LP_neutrino} and~\cite{nikakhtar2021BAO_Laguerre}).~However, in practice, provided that the fits are good, neither the $r_{\rm LP}$ estimates nor their error bars depend strongly on which functional form is fit.~Our analysis of the covariance matrix allows us to provide a rather general estimate of the expected precision that is not tied to a particular functional form.~It also allows us to address a closely related question.~In principle, the inflection point $r_{\rm infl}$, the scale on which $d^2\xi/dr^2 = 0$, could also be used as a standard rod \citep{anselmi2016LP_BAO}.~Previous work has suggested that it is less robust than the LP \citep{anselmi2018LP_mock_catalogs, nikakhtar2021BAO_Laguerre_mock_catalogues}; our analysis provides some insight into why this is so.

This paper is organized as follows.  Section~\ref{sec:methods} describes how the eigenvalues and eigenvectors of the TPCF covariance matrix change as the shot-noise increases, and then use this to provide a simple estimate of the error on the LP.~Section~\ref{sec:binning} shows how our results depend on the binning. Section~\ref{sec:conclusion} summarizes our conclusions.

Although we focus on the Linear Point scale, the scale $r_0$ where the pair correlation function crosses zero, $\xi(r_0)=0$, has also been proposed for use as a standard ruler \cite{Prada2011}.~Appendix \ref{sec:r0} shows the results of applying our analysis to $r_0$.

\section{Methods and Results}\label{sec:methods}

Because the neighboring bins of the TPCF amplitudes are correlated, the covariance matrix of the bin counts is not diagonal.~Here below, we describe in detail how we use the structure of the covariance matrix to estimate realistic error bars on the BAO distance scale. 

Where necessary to illustrate our results, we will use a comoving volume of $5 \times (1.024)^3$ $h^{-3}$Gpc$^3$ in a flat $\Lambda$CDM model with $(\Omega_{m,0},\Omega_{b,0}) = (0.281,0.046)$, and $(h,n_s,\sigma_8) = (0.697,0.971,0.842)$ as in \cite{BAO_LP_MG}.~The associated linear theory values of $r_{\rm LP}$ and $r_{\rm infl}$ 
for the dark matter are $97.154~h^{-1}$Mpc and $97.635~h^{-1}$Mpc, respectively.~For easy comparison with Refs.\cite{nikakhtar2021BAO_Laguerre,BAO_LP_MG}, we focus on biased tracers at $z=0.5$ (also denoted as $z = 0.5057$ following \cite{BAO_LP_MG}).~Although we explore other combinations of number density and clustering strength, our fiducial choice has $\bar{n}=3.2\times 10^{-4}/(h^{-1}{\rm Mpc})^3$ and linear bias factor $b\approx 1.97$, which is similar to the Baryon Oscillation Spectroscopic Survey (BOSS) and the Dark Energy Spectroscopic Instrument (DESI) survey \cite{sdss_dr12,DESI2024III_BAO}.~The combination $\bar{n}P(k_{\rm max})$, where $k_{\rm max}$ is the scale on which $P(k) = b^2\,P_{\rm Lin}(k)$ is maximum, is sometimes used as a crude measure of whether the BAO clustering signal is dominated by shot-noise.~Our fiducial choice has $\bar{n}P(k_{\rm max})\sim 5$; shot-noise dominates for values smaller than unity. While we provide our formalism in terms of the real-space correlation function, we use the redshift-space monopole in our figures to show that our methodology is valid even under redshift-space distortions.~This makes $b_{\rm eff} = \sqrt{(b^2 + 2bf/3 + f^2/5)}$ with $f = d\ln D/d\ln a$ being the linear theory growth rate \cite{kaiser1987rsd}.  With $b\sim 1.97$ at $z\sim 0.5$, $b_{\rm eff}\sim 2.23$.

\subsection{Gauss-Poisson approximation to covariance matrix}
We begin with the two-point correlation function, which is related to the power spectrum $P(k)$ by
\begin{equation}
    \xi(s) \equiv\int \frac{dk}{k}\frac{k^3P(k)}{2\pi^2}\,j_0(ks).
     \label{eq:xis}
\end{equation}
A crude model for nonlinear evolution sets 
 $P(k) = P_{\rm Lin}(k)\,{\rm e}^{-k^2\Sigma^2}$ 
\cite[e.g.][]{rpt2008}, so the nonlinear correlation function is a `smeared' version of the linear one.~At $z=0.5$ in our fiducial cosmology, $\Sigma \approx 4.7~h^{-1}$Mpc for the real-space TPCF; it is slightly larger, $\Sigma_{\rm eff} \approx 6.3~h^{-1}$Mpc, for the biased redshift-space monopole \cite{nikakhtar2021BAO_Laguerre_mock_catalogues,paranjape2023}.

If the correlation function is estimated by counting pairs separated by $s\pm\Delta s/2$ in a discrete set of $N_{\rm tot}$ particles distributed in a volume $V_{s}$, then the TPCF covariance matrix described by the `Gauss-Poisson' approximation is given by \cite{smith2008peak,grieb2016gaussian_covariance,parimbelli2021LP_neutrino}: 
\begin{equation}
    C_{ij} = \frac{2}{V_{s}}\int_{0}^{\infty} \frac{dk~k^2}{2\pi^2} \bar{j_{0}}(ks_i)\bar{j_{0}}(ks_j) \big[ P(k) + \frac{1}{\bar{n}}\big]^2 
\label{eq:TPCF_cov_mat_xi0}
\end{equation} 

\noindent where $\bar{n}\equiv N_{\rm tot}/V_{s}$ is the survey number density and $\bar{j_{0}}$ is the bin-averaged spherical Bessel function: 
\begin{equation}
    \bar{j_{0}}(ks_i) \equiv \frac{4\pi}{V_{s_i}}\int_{s_i-\Delta s/2}^{s_i+\Delta s/2} s^2 j_{0}(ks) ds,
\end{equation}
\noindent with the volume $V_{s_i} = 4\pi/3 (s_{i,\rm{max}}^3 - s_{i,\rm{min}}^3) $, $s_i$ being the midpoint of bin $i$, and $\Delta s$ being the bin-size.~The shot-noise only term proportional to $1/\bar{n}^2$ only contributes when $i = j$, i.e., to the error bar in a single bin.~The other two terms describe the covariance between bins, and come from `cosmic variance.'~This covariance must be accounted for when estimating the uncertainty on the BAO scale. 

\begin{figure}[t]
\includegraphics[width=0.49\textwidth]{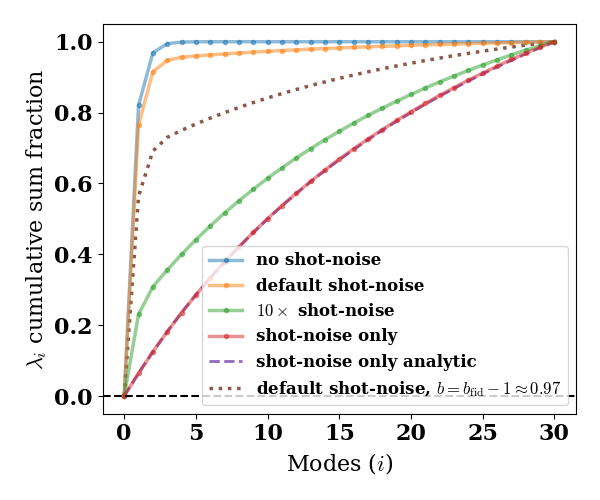}
\caption{\label{fig:eigvals_by_shot}Fractional contribution of the sorted eigenvalues $\lambda_j$ to the total variance, $\sum_1^{i}{\lambda_j}/\sum_1^{30}{\lambda_j}$, for a range of choices of the ratio of `signal' ($P(k)$ amplitude) to `noise' ($1/\bar{n}$).~Fiducial and idealized no shot-noise cases are very similar; more than 95\% of the total variance comes from the first 3 modes.~More modes matter when the noise dominates.~The smooth dashed curve shows our analytic expression for the pure noise case: $(2/\bar{n}V)/(\bar{n}V_i)$.~Dotted curve is for $b\to b-1$ but fiducial shot-noise.}
\end{figure}

\begin{figure*}
\includegraphics[width=0.98\textwidth]{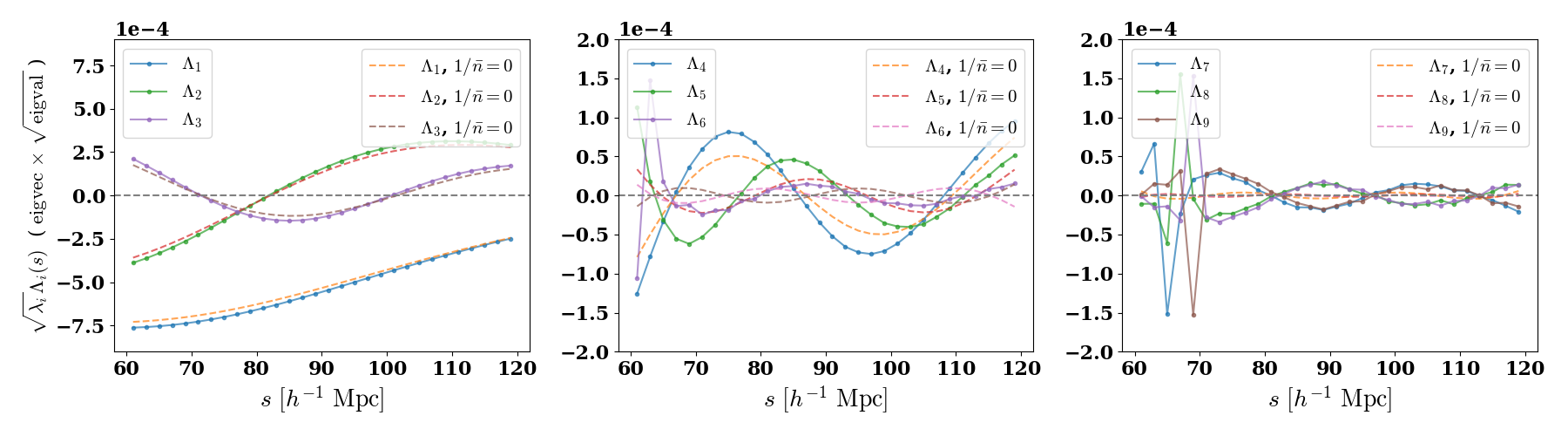}
\caption{\label{fig:eigvals_eigvecs_vs_no_shot}The first 9 eigenvectors $\sqrt{\lambda_i} \Lambda_i(s)$ at $z = 0.5057$, for default shot-noise $\bar{n} = 3.2 \times 10^{-4}/(h^{-1}{\rm Mpc})^3$ (solid) and no shot-noise (dashed).~The first 3 modes (left hand panel), which contribute most to the total variance, are much less sensitive to shot-noise. 
Higher order modes (right hand panel, note the factor of 4 difference in the $y$-axis scale) are essentially zero when there is no shot-noise, and resemble delta-functions which peak at different scales, followed by oscillations, when shot-noise is present.} 
\end{figure*}

\subsection{Eigen-decomposition of the covariance matrix}\label{sec:covariance_eigendecomposition}
To illustrate our results, we now use Eq.~\ref{eq:TPCF_cov_mat_xi0} to generate $C_{ij}$, for 30 non-overlapping bins of $\Delta r = 2~h^{-1}$Mpc, running from $[60-120]~h^{-1}$Mpc, with the fiducial values of background cosmology, redshift, survey volume, biased tracer number density, and clustering strength mentioned at the start of this section.    

Next, we diagonalize $C_{ij}$.~The eigenvectors, which we denote $\Lambda_i (s)$, provide a set of orthogonal shape functions, whose relative importance is set by the eigenvalues $\lambda_i$.~Before we consider the interplay of $P$ and $\bar{n}$ in determining these eigen-modes, note that the survey volume $V_s$ only appears as an overall scaling.~It scales the eigenvalues up and down but keeps their ratios fixed, and does not affect the eigen-shapes.~That said, $V_s$ is important because it does {\em not} enter in the definition of $\xi$ itself.~Therefore, larger $V_s$ means that the eigen-modes will have smaller amplitudes compared to $\xi$.  This will be important below.

Before we look at the shapes, Fig.~\ref{fig:eigvals_by_shot} shows the fractional contribution of the eigenvalues $\lambda_i$ (ordered from largest to smallest) to the total variance.~The various curves show different choices for the relative contributions of `signal' $P(k)$ and `shot-noise' $1/\bar{n}$.  

The lowermost curve is for the pure shot-noise limit (we have set $P(k)=0$).~This case is analytic:  $C_{ii}$ is diagonal, with entries $(2/\bar{n}V)/(\bar{n}V_i)$, where $V_i$ is the volume of the $i$th bin.~For bins of width $\Delta r = r_{i+1} - r_{i}$ that are equally spaced with spacing $\epsilon\Delta r$ (typically $\epsilon=1$ but we will see later why the more general case is interesting) where $r_i$ is the lower bound of the $i$th bin, 
$V_i = (4\pi/3) (\Delta r)^3 [(j\epsilon - (\epsilon - 1)/2)^3 - (j\epsilon - 1 - (\epsilon - 1)/2)^3]$ where $j = i + 60/(\epsilon\Delta r)$ and the bin centers are given by $\frac{(r_i + r_{i+1})}{2} =60 + (i-1) \epsilon \Delta r + \epsilon\Delta r/2$ since our bins start at $60~h^{-1}$Mpc.~Note that all eigenvectors matter.~These eigenvectors are delta functions, one for each bin, centered on the middle of the bin.

In contrast, the uppermost curve shows the case in which $1/\bar{n}=0$:~here $C_{ij}$ is completely determined by the $P^2$ term in Eq.~\ref{eq:TPCF_cov_mat_xi0}.~Notice that now the variance is dominated by just a few eigenvalues/eigenvectors.~The intermediate curves show different choices for the $1/\bar{n}$.~The `fiducial' choice ($\bar{n}P(k_{\rm max})\approx 5$) is very similar to the no-noise case.~However, as the noise increases, more modes begin to matter.\footnote{The cross term in $C_{ij}$ is also analytic:  it is a smoothed version of $\xi$, with smoothing depending on scales $i$ and $j$, but the expression is lengthy so we have not written it explicitly here.}

Fig.~\ref{fig:eigvals_eigvecs_vs_no_shot} shows the corresponding eigenvectors.~Notice that the $n$th eigenvector has $n-1$ zero-crossings, at least for the first few $n$.~It is striking that the mode with 4 zero-crossings divides the 60-120$~h^{-1}$Mpc range up into patches that are approximately the size of the BAO feature itself.~Presumably this is because the same $P(k)$ appears in both $\xi$ and $C_{ij}$.~Dashed curves are for the case of no shot-noise, and solid curves have the fiducial shot-noise.~Recall that these are the cases that are dominated by the first few eigenvalues, and the associated eigenvectors are very similar and very smooth.~The smoothness is consistent with the expectation that terms contributing to cosmic variance should be smooth functions of scale.~However, the similarity is particularly interesting here:~it suggests that the eigenvectors for the no-noise case remain interesting even in the presence of fiducial noise.  

The other higher-order eigenvectors, which contribute little to the total variance, are more strongly modified by the presence of shot-noise.~To gain some insight, recall that the pure shot-noise eigenvectors would be a set of delta functions, each centered on a bin.~But, when $P$ is significant, these delta functions are now approximately rotated into the basis in which the $P^2$ term is diagonal.~This mixes the delta functions, and is why these higher order modes display oscillations.~We will exploit this relatively clean separation into cosmic variance vs. shot-noise dominated eigen-modes in the next section.

\subsection{Eigen-decomposition of correlation function realizations}

We can write one realization $\Xi$ of the real-space TPCF as: 
\begin{equation}
    \Xi(s) = \xi(s) + \sum_i g_i \Lambda_i(s), 
\label{eq:Xi_realization}
\end{equation}
where $\xi(s)$ is given by Eq.~\ref{eq:xis}, the $g_i$'s are independent Gaussian random variates with variance $\lambda_i$ and mean zero, while $\Lambda_i(s)$ are the eigenvectors of $C_{ij}$.~Hence, the terms other than $\xi(s)$ represent the (correlated) scatter around the mean.

\begin{figure}
\includegraphics[width=0.49\textwidth]{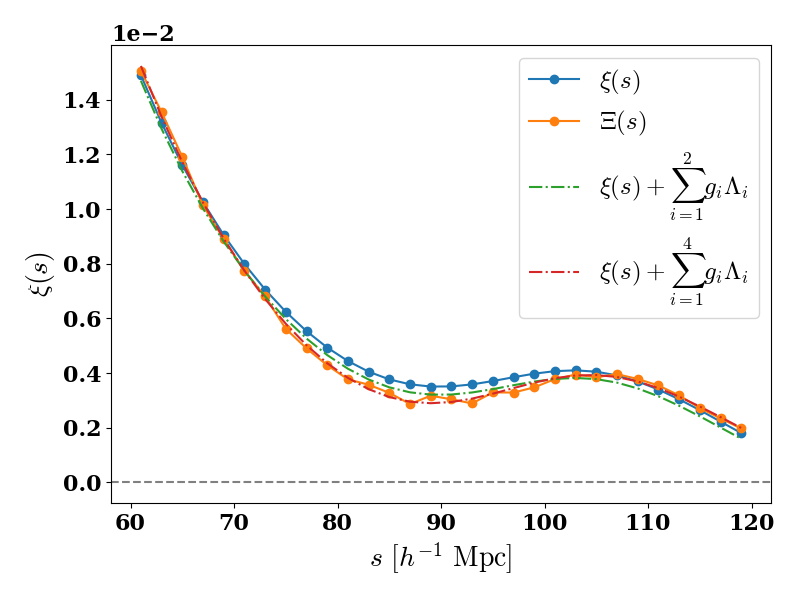}
\caption{\label{fig:xi0_by_mode}$\xi(s)$, $\Xi(s)$, and Eq.~\ref{eq:Xi_realization} with the first 2 and 4 modes.~For a survey volume smaller than our fiducial $V_s$ of $\sim 5~(h^{-1} \rm{Gpc})^3$, the deviations of the other 3 curves from the blue $\xi(s)$ curve would be larger. 
}
\end{figure}

\begin{figure}
\includegraphics[width=0.49\textwidth]{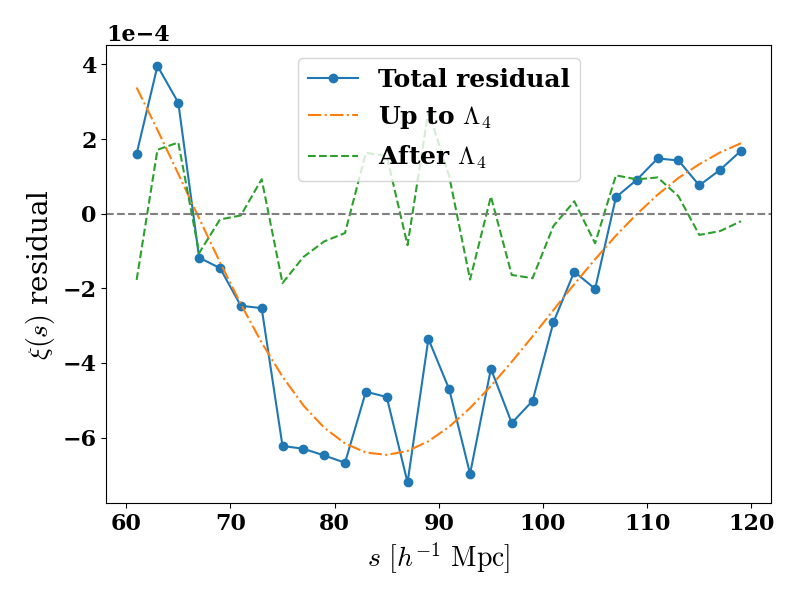}
\caption{\label{fig:residual}Total residual $\Xi(s)- \xi(s)$ for one TPCF realization (symbols), and the contributions to it from modes 1 to 4 (dot-dashed), and from 5 onwards (dashed).}
\end{figure}

The symbols in Fig.~\ref{fig:xi0_by_mode} show how the shape of $\Xi$ changes as more modes are added to $\xi$, for one realization where we have assumed the fiducial noise and bias.~The changes are relatively mild because $V_s$ is sufficiently large that the $\lambda_i$ are small.~To highlight the differences as more modes are added, the symbols in Fig.~\ref{fig:residual} show the total residual from the mean, $\Xi(s)- \xi(s) = \sum_i g_i \Lambda_i(s)$, for this same realization.~The other curves show the contribution from modes 1 to 4 ($\sum_{i = 1}^{i = 4} g_i\Lambda_i(s)$), and from 5 onwards ($\sum_{i = 5}^{i = 30} g_i \Lambda_i(s)$).~Clearly, the first 4 modes capture most of the residual, including the small change in shape, while the sum of modes 5 and onwards is mostly uncorrelated noise (small amplitude oscillations around zero).~This just illustrates what Fig.~\ref{fig:eigvals_by_shot} showed:~the higher order modes are not particularly important.  

\subsection{Truncation of modes and basis functions for fitting the TPCF}
Because we see that the sum of the first 4 modes captures most of the residual, while the remaining modes are mostly `noise,' it is interesting to consider restricting the sum in Eq.~\ref{eq:Xi_realization} to include only the first $\sim 4$ modes.~Evidently, this removes the shot-noise dominated fluctuations from the realization of $\Xi$, leaving a smoother curve.~In essence, this is the smooth curve one is after when `fitting' the correlation.~This is a non-trivial statement, since $\xi$ $+$ the first few eigenvectors, while smoother than the full $\Xi$, will not generically have the same shape as $\xi$.~Nevertheless, since typical datasets were designed (i.e.~$V_s$ is large enough) so that cosmic variance does not dominate, the amplitude of these `cosmic variance' modes is small compared to the amplitude of $\xi$ itself, so the correction to the shape is small (c.f. Fig.~\ref{fig:xi0_by_mode}).  

In the same vein, suppose one is interested in derivatives of the correlation function.~Although 
\begin{equation}
    \Xi^{\prime} = \xi^{\prime} + \sum_{i} g_i \Lambda_i^{\prime} ,
     \label{eq:dXids}
\end{equation}
if we include all 30 terms in the sum, then this will be like differentiating a single measurement of the correlation function.  However, it is well known that one should {\em not} differentiate a noisy measurement; rather, one should first fit a smooth functional form to the measurement and then differentiate the fit.  In the present context, our model for this procedure is to assert that one is {\em not} interested in $d\Xi/ds$ when the sum includes all 30 terms; rather, one should only include the first 4 (really, the ones which account for, say, 90\% of the variance). 

The virtue of this point of view is that this shape is clearly determined {\em solely} by the shape of $P(k)$; in particular, it makes no reference to the set of basis functions which one wishes to fit to $\Xi$ (simple polynomials?~Laguerre functions?~etc.).~This is attractive, since a reasonable concern is whether the set of basis functions which worked for one underlying $P(k)$ will also work for another (polynomials for one, Laguerres for another?). Here, the point is that one should think of the eigenvectors $\Lambda_i$ as being the most appropriate set of basis functions, since these are clearly determined by the shape of $P(k)$.

\subsection{Model-independent error estimates on BAO distance scale}
We will now use this insight to discuss how one might quantify uncertainties on estimates of the BAO scales $r_{\rm{LP}}$, $r_{\rm{Dip}}$ and $r_{\rm{Peak}}$.~We also consider, $r_{\rm{infl}}$, the scale on which the second derivative vanishes, as an alternative to $r_{\rm LP}$.  

We start with Eq.~\ref{eq:dXids}, but restrict the sum to the first few terms (the ones which account for, say, 90\% of the variance).~Next, note that the scale $s_{\rm{max}}$ where $\Xi^{\prime} = 0$ is not necessarily the same as $r_{\rm{max}}$, where $\xi^{\prime} = 0$. Assuming $\Delta_{\rm{max}} \equiv s_{\rm{max}} - r_{\rm{max}} \ll 1$, we have when $\Xi^{\prime} = 0$, 

\begin{equation}
    \begin{split}
    0 &= \xi^{\prime}(s_{\rm{max}}) + \sum_{i} g_i \Lambda_i^{\prime}(s_{\rm{max}}) \\
    & \approx \xi^{\prime} (r_{\rm{max}}) + \Delta_{\rm{max}} \xi^{\prime \prime} (r_{\rm{max}}) + \sum_{i} g_i \Lambda_i^{\prime}(s_{\rm{max}}) \\
    & \approx \Delta_{\rm{max}} \xi^{\prime \prime} (r_{\rm{max}}) + \sum_{i} g_{i} \big[ \Lambda_i^{\prime}(r_{\rm{max}})  + \Delta_{\rm{max}} \Lambda_i^{\prime \prime}(r_{\rm{max}})  \big], 
    \end{split}
\label{eq:Delta_derivation}
\end{equation}

\noindent where we have used that $\xi^{\prime} = 0$ at $r_{\rm{max}}$.~Isolating $\Delta_{\rm{max}}$ yields

\begin{equation}
    \frac{\Delta_{\rm{max}}}{r_{\rm max}} = - \frac{\sum_{i} g_i \,(\Lambda_i^{\prime}/r_{\rm{max}})/\xi^{\prime\prime}}{1 +  \sum_{i} g_i (\Lambda_i^{\prime \prime}/\xi^{\prime \prime}) }.
\label{eq:Delta_max_realization}
\end{equation}

\noindent In practice, on the peak and dip scales, $(\Lambda_i^{\prime \prime}/\xi^{\prime \prime})\le 0.1$ or so.  Since $\langle g_i^2 \rangle = \lambda_i$ is also small, we can neglect the term in the denominator and approximate 

\begin{equation}
    \bigg\langle \frac{\Delta_{\rm{max}}^2}{r_{\rm{max}}^2} \bigg\rangle \approx \sum_i \lambda_i\, \bigg( \frac{ \Lambda_i^{\prime}(r_{\rm{max}})}{r_{\rm{max}}\xi^{\prime \prime}(r_{\rm{max}}) } \bigg)^2  .
\label{eq:Delta_max}
\end{equation}

\noindent Eq.~\ref{eq:Delta_max} shows that the root-mean-square (RMS) of $\Delta_{\rm{max}}$ increases as $|\xi^{\prime \prime}|$ decreases.~Since the nonlinear TPCF is more smeared (i.e. less curved at the peak and dip scales) than the linear theory TPCF (recall discussion of Eq.~\ref{eq:xis}), we expect the uncertainty in the peak scale to be larger in the evolved field (at lower redshifts).  We return to this point in the next subsection. 

The same logic can be applied to the dip scale, so 

\begin{equation}
    \bigg\langle \frac{\Delta_{\rm{min}}^2}{r_{\rm{min}}^2} \bigg\rangle \approx \sum_i \lambda_i\, \bigg( \frac{ \Lambda_i^{\prime}(r_{\rm{min}})}{r_{\rm{min}}\xi^{\prime \prime}(r_{\rm{min}}) } \bigg)^2 .
\label{eq:Delta_min}
\end{equation}
Hence, just as for $\Delta_{\rm{min}}$, the RMS of $\Delta_{\rm{min}}$ in the evolved TPCF is larger than the linear theory value. 

Finally, for the error on the LP scale, 

\begin{equation}
    \Delta_{\rm{LP}} \equiv \frac{\Delta_{\rm{max}} + \Delta_{\rm{min}}}{2},
\end{equation}

\noindent we have:

\begin{equation}
    \langle \Delta_{\rm{LP}}^2 \rangle \approx \frac{\langle \Delta_{\rm{max}}^2 \rangle}{4} + \frac{\langle \Delta_{\rm{min}}^2 \rangle}{4} + \sum_i \frac{\lambda_i}{2} \frac{ \Lambda_i^{\prime}(r_{\rm{max}})}{\xi^{\prime \prime}(r_{\rm{max}}) } \frac{ \Lambda_i^{\prime}(r_{\rm{min}})}{\xi^{\prime \prime}(r_{\rm{min}}) }.
\label{eq:Delta_LP}
\end{equation}

\noindent Because $\xi^{\prime \prime}(r_{\rm{max}})$ and $\xi^{\prime \prime}(r_{\rm{min}})$ have opposite signs (by definition), the variance of $\Delta_{\rm{LP}}$ is smaller than either $\Delta_{\rm{Dip}}$ or $\Delta_{\rm{Peak}}$. This demonstrates why the LP is a more precise probe than either $r_{\rm{Dip}}$ and $r_{\rm{Peak}}$. 

Similarly for the inflection point:~if $r_{\rm infl}$ is the scale where $\xi^{\prime\prime} = 0$, and $s_{\rm infl}$ is the scale where $\Xi^{\prime\prime}=0$ (when the sum which defines $\Xi$ is truncated to only include the eigenvalues whose eigenvectors contribute $\sim 90\%$ of the variance), we would set 
\begin{equation}
    \begin{split}
    0 &= \xi^{\prime\prime}(s_{\rm{infl}}) + \sum_{i} g_i \Lambda_i^{\prime\prime}(s_{\rm{infl}}) \\
    & \approx \Delta_{\rm{infl}} \xi^{\prime\prime \prime} (r_{\rm{infl}}) + \sum_{i} g_{i} \big[ \Lambda_i^{\prime\prime}(r_{\rm{infl}})  + \Delta_{\rm{infl}} \Lambda_i^{\prime\prime \prime}(r_{\rm{infl}})  \big], 
    \end{split}
\end{equation}
where $\Delta_{\rm infl} = s_{\rm infl} - r_{\rm infl}$.  Thus,
\begin{equation}
    \frac{\Delta_{\rm infl}}{r_{\rm infl}} = -
    \frac{\sum_{i} g_{i} (\Lambda_i^{\prime\prime}/r_{\rm infl})/\xi^{\prime\prime\prime}}{1 + \sum_{i} g_{i} (\Lambda_i^{\prime\prime\prime}/\xi^{\prime\prime\prime})}
 \approx -\sum_{i} g_{i} \,\frac{\Lambda_i^{\prime\prime}}{r_{\rm infl}\,\xi^{\prime\prime\prime}}.
\label{eq:Delta_infl}
\end{equation}
so
\begin{equation}
    \bigg\langle \frac{\Delta_{\rm{infl}}^2}{r_{\rm{infl}}^2} \bigg\rangle \approx \sum_i \lambda_i\, \bigg( \frac{ \Lambda_i^{\prime\prime}(r_{\rm{infl}})}{r_{\rm{infl}}\,\xi^{\prime\prime\prime}(r_{\rm{infl}}) } \bigg)^2 .
\label{eq:Delta_infl_approx}
\end{equation}

\subsection{Comparison with standard method}

\begin{figure}
\includegraphics[width=0.49\textwidth]{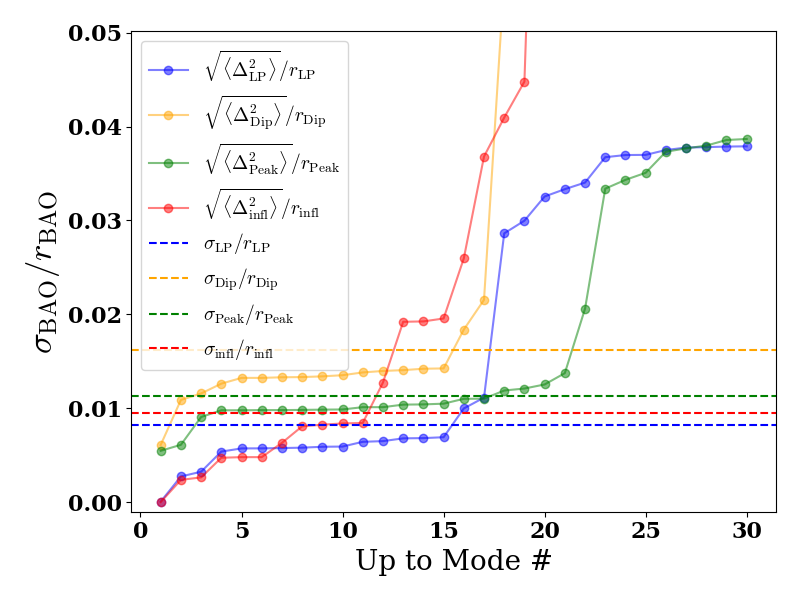}
\caption{\label{fig:sigma_BAO_estimates}Fractional uncertainties on the dip, peak, inflection point and LP scales (square root of 
Eqs.~\ref{eq:Delta_min},~\ref{eq:Delta_max},~\ref{eq:Delta_infl},~\ref{eq:Delta_LP}), as a function of the number of eigen-modes that are included (symbols).~Horizontal dashed lines show the corresponding uncertainties estimated in the standard way (see main text).}
\end{figure}

To see how well this works, we first estimate the four scales (peak, dip, LP and inflection point) in the standard way \cite[e.g.][]{anselmi2016LP_BAO, anselmi2019LP_fitting, parimbelli2021LP_neutrino, nikakhtar2021BAO_Laguerre, paranjape2023, BAO_LP_MG}:~we made 100 mock realizations of the measurement (Eq.~\ref{eq:Xi_realization}), fitted a 7th order polynomial to each, and estimated the various scales by differentiating the fit.~The RMS scatter of each scale satisfies $\sigma_{\rm dip} > \sigma_{\rm peak} > \sigma_{\rm infl} > \sigma_{\rm LP}$:~the LP is the most precise, followed by the inflection point, and etc.~This is consistent with previous work 
(e.g.~Tables~4 to 7 of \cite{BAO_LP_MG}).~The actual values are shown as the horizontal dashed lines in Fig.~\ref{fig:sigma_BAO_estimates}.~Previous work has shown that it does not matter if one fits 7th order generalized Laguerre functions instead \citep{nikakhtar2021BAO_Laguerre}.

Now we turn to our eigen-mode-based estimates.  
Eqs.~\ref{eq:Delta_min},~\ref{eq:Delta_max},~\ref{eq:Delta_infl}, ~\ref{eq:Delta_LP} show that these depend on the number of modes that are included in the relevant sums.~The symbols in Fig.~\ref{fig:sigma_BAO_estimates} show how these estimates increase as more modes are added.~We previously argued that one should not include the higher order modes (because one should not differentiate noisy data); these are the ones for which the error estimate starts to diverge.~The plateau at intermediate values indicates that the error estimate is not very sensitive to exactly how many modes are included, provided we have enough modes, and are not including the ones which are dominated by shot-noise. (This plateau is not an artifact produced by approximation Eq.~\ref{eq:Delta_max}; it is present even if we use Eq.~\ref{eq:Delta_max_realization}.)~In effect, this plateau provides an objective measure of how many modes should be included to accurately model the scatter between realizations, analogous to how Fig.~\ref{fig:eigvals_by_shot} provides an objective way of deciding which modes are most important for a single realization.~Indeed, previous work has noted that the inflection point is slightly less robust than the LP:  here, this is indicated by the fact that it has a shorter plateau. 

\begin{figure}
\includegraphics[width=0.49\textwidth]{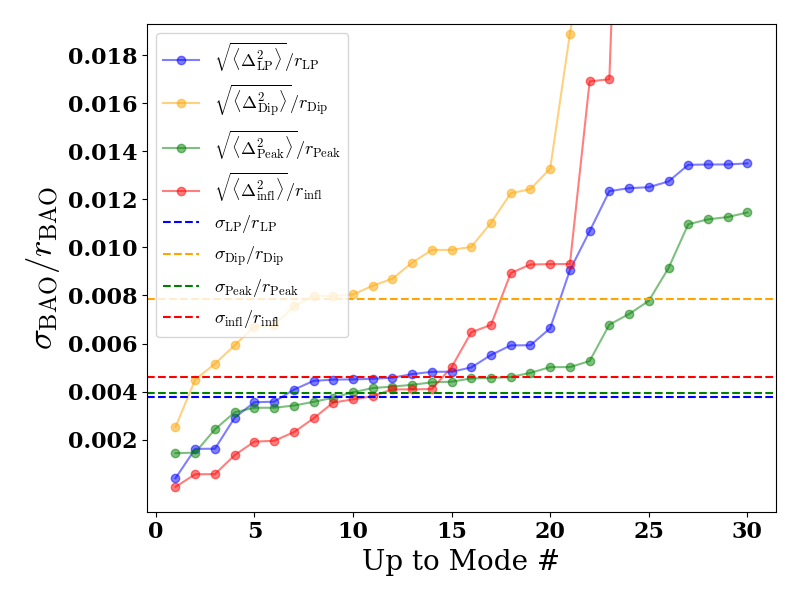}
\caption{\label{fig:sigma_BAO_estimates_linear}Same as Fig.~\ref{fig:sigma_BAO_estimates}, but with $\Sigma=0$ (no smearing), for both $\xi_0(s)$ and the covariance matrix.~Comparison with Fig.~\ref{fig:sigma_BAO_estimates} (note the difference in $y$-axis range) shows the potential gains in precision which come from working with the reconstructed BAO signal. 
}
\end{figure}

It is reassuring that, not only do the plateau values reproduce the qualitative trends shown by the standard method, they are within 80\% of the fitted RMS values in all cases.~Some of this discrepancy arises from the assumption that the scale which the standard method identifies as being the peak, say $t_{\rm max}$  ($t$ for template), may differ slightly from $s_{\rm max}$ (our eigen-mode based estimate of the peak scale).~As a result, the RMS of $t_{\rm max}-r_{\rm max}$ differs from the RMS of $s_{\rm max}-r_{\rm max}$; evidently, this difference is small.

Recall that the standard method results do not depend on the functional form that was fit to the binned TPCF.~In effect, our analysis shows why:~the low order eigenvectors $\Lambda_i$ represent the covariance around any good fit to the measurements which is not `fitting the noise.'  


\interfootnotelinepenalty=10000

We noted that, because the BAO feature in $\xi$ becomes more smeared at late times, $\xi''$ decreases, so we expect the uncertainty on the LP scale to increase (c.f.~Eqs.~\ref{eq:Delta_max}--\ref{eq:Delta_LP}).~Fig.~\ref{fig:sigma_BAO_estimates_linear} tests this:~it shows the same comparison as in Fig.~\ref{fig:sigma_BAO_estimates}, but now with $P(k) \propto P_{\rm Lin}(k)$ (i.e. $\Sigma=0$, no smearing) for determining both $\xi_0(s)$ and $C_{ij}$ when using 
Eq.~\ref{eq:Xi_realization} to produce 100 realizations of $\Xi(s)$.~Setting $\Sigma\to 0$ changes the BAO feature in $\xi_0$ dramatically, and $C_{ij}$ less so:~e.g.~the total variance is about 15\% larger, and eigenvalues 3 to 7 contribute considerably more to the total variance compared to when $\Sigma\ne 0$. The change in $\xi_0$ means that a 7th order polynomial is no longer a good fit, so, for the `standard' analysis, we used a 9th order polynomial.~Comparison of the dashed lines here with those in the previous figure shows that the RMS is decreased by a factor of about 2 to 3.~This is especially true for the peak which is most affected by the smearing.\footnote{It may seem surprising that, in contrast to the evolved field, the peak and LP scales in the linear field are measured with similar precision.~This is mainly a signal-to-noise issue.~Recall that the original reason for working with the LP was not its precision, but its robustness to evolution/smearing \cite{anselmi2016LP_BAO}.}~Notice that this decrease is reproduced by our eigen-mode estimates.~The onset of the plateau is delayed from about mode 4 to about mode 7 or so, since now it is the first 7 modes which contribute to most of the variance.~This decrease demonstrates the potential gains which come from using the reconstructed TPCF rather than the smeared one to measure the BAO scales:~for the LP, this results in a precision of better than 0.4\% as opposed to 0.8\%.  Our analysis has allowed us to estimate this improvement without having to run simulations. 

Some reconstruction methods move the observed biased tracers back to (an estimate of) their initial positions \cite{OTbao, OTrsd, OTmt}.~The TPCF is then measured using these reconstructed positions.~In this case, the number density is unchanged but $b\to b-1$:~typically, the BAO feature is sharper, but the amplitude is smaller \cite{OTbao}.~Our methodology allows an estimate of the precision of the LP distance scale in this reconstructed signal as follows.~Since $\bar{n}(b-1)^2P_{\rm Lin}$ is smaller than $\bar{n}b^2P_{\rm Lin}$, the dotted curve in Fig.~\ref{fig:eigvals_by_shot} suggests that more modes will be needed before we converge to a plateau as in Fig.~\ref{fig:sigma_BAO_estimates}.~Although we do not show it here, we have checked that this is indeed the case. In addition, the precision of the reconstructed feature is slightly {\em less} precise, less constraining, than the original measurement.~I.e., reconstruction yields no significant gain in precision (of course, it does reduce the bias in the mean value, bringing the LP closer to its linear theory value).~To realize the potential for increased precision shown in Fig.~\ref{fig:sigma_BAO_estimates_linear}, one must combine reconstructed fields, as discussed in \cite{OTmt}.

We conclude that our methodology is an efficient way of determining accurate uncertainties on the Linear Point estimate of the BAO distance scale.~In particular, since our analysis suggests that, for reasonable/realistic values of the shot-noise, the relevant eigenvalues and eigenvectors are entirely determined by the ratio of $P^2(k)$ and the survey volume $V_s$, they scale as $(b\sigma_8)^2/\sqrt{V_s}$.~Our curves assumed $(b,\sigma_8,V_s)=(1.97,0.84,5~h^{-3}{\rm Gpc}^{3})$, so for other values, the fractional error is given by scaling the numbers in our Fig.~\ref{fig:sigma_BAO_estimates} by $(b_{\rm eff}/2.23)^2(\sigma_8/0.84)^2\sqrt{5/V_s}$.~This is not quite right, because it ignores the fact that the smearing also depends on $\sigma_8$, but it is a useful first guess.~Additionally, we note that the number of eigen-modes that captures a sufficient amount of information depends on $bP(k)$, $V_{s}$, as well as $\bar{n}$.~There are two features in our analysis that show this number:~(1) the change of the slope in Fig. \ref{fig:eigvals_by_mode} where we plot the eigenvalues by mode and (2) the plateaus in Figs.~\ref{fig:sigma_BAO_estimates},  \ref{fig:sigma_BAO_estimates_linear}, and \ref{fig:sigma_r0} that show $\sigma_{\rm BAO}$ (which is what we really care about).~Although we do not provide an explicit expression here, both features suggest that the first 5 eigen-modes (which contributes to $\sim 95\%$ of the total covariance) are sufficient for our choice of $(b, V_{s}, \bar{n})$, which are reasonable values for current and next-generation surveys.~While our fiducial choice of binning is 30 non-overlapping bins of $2~h^{-1}$Mpc, our results are rather insensitive to binning - with 5 eigen-modes being sufficient for 15 bins of $4~h^{-1}$Mpc, as well as 10 bins of $6~h^{-1}$Mpc, the reason for which we will see in Section \ref{sec:binning}.

\section{Dependence of eigenvectors on binning}\label{sec:binning}
The previous section considered the structure of the covariance matrix of a binned estimate of the TPCF.  In that analysis, the original bin-size (and shape) was fixed.~How is the analysis modified if we change the binning?  

In what follows, we first show that, in general, the eigenvectors of the binned covariance matrix are not simply binned versions of the original eigenvectors.~Nevertheless, the first few eigenvectors are unchanged by the variations in binning, permitted by the requirement that one be able to detect the BAO feature in the first place.  

\subsection{Analytic analysis}
Let $\mathbf{x}$ denote our list of bins, $\mathbf{y}$ the list of measured bin amplitudes, $\mathbf{C}\equiv \langle \mathbf{yy}^{\rm T}\rangle$ the covariance matrix of the measurements, and $\lambda_i$ and $\mathbf{\Lambda}_i$ its eigenvalues and eigenvectors.~With some abuse of notation, let $\bm{\lambda}$ denote the diagonal matrix with the eigenvalues along the diagonal.  If $\mathbf{V}$ is a square matrix whose columns are the eigenvectors $\mathbf{\Lambda}$, then 
\begin{equation}
  \mathbf{C} = \mathbf{V}\bm{\lambda}\mathbf{V}^{-1}
             = \mathbf{V}\bm{\lambda}\mathbf{V}^{\rm T} ,
\end{equation}
where the final equality follows because $\mathbf{C}$ is real and symmetric.

Suppose we bin so that $\mathbf{y}_B\equiv \mathbf{By}$, with $\mathbf{B}$ a square matrix, each side having the same dimension as $\mathbf{y}$.  E.g.,
\begin{equation}
   \label{eq:matrixB}
   \mathbf{B} = \frac{1}{3}
  \begin{pmatrix}
        1 & 1 & 0 & 0 & \cdots & 0 & 0 & 0 \\ 
        1 & 1 & 1 & 0 & \cdots & 0 & 0 & 0 \\ 
        0 & 1 & 1 & 1 & \cdots & 0 & 0 & 0 \\ 
   \vdots & \vdots & \vdots & \vdots & \ddots & \vdots & \vdots & \vdots \\
        0 & 0 & 0 & 0 & \cdots & 1 & 1 & 1\\
        0 & 0 & 0 & 0 & \cdots & 0 & 1 & 1 \\ 
  \end{pmatrix} 
\end{equation}
would correspond to averaging the $y$ values in the bins on either side of each $x$-bin.~Note that, when this is done, one typically works with a sparser set of $\mathbf{x}$ values, so as to not `double-count.'~The analysis below is more transparent when
$\mathbf{x}_{\rm B}$ and $\mathbf{x}$, and hence 
$\mathbf{y}_{\rm B}$ and $\mathbf{y}$, have the same length.  

If $\mathbf{C}_{\rm B}$ denotes the covariance matrix of the binned measurements, then
\begin{align}
  \label{eq:binnedC}
  \mathbf{C}_{\rm B} &\equiv \langle \mathbf{By\, (By)}^{\rm T}\rangle
  = \mathbf{B} \mathbf{C} \mathbf{B}^{\rm T}\nonumber\\
  &= \mathbf{B}\, (\mathbf{V}\bm{\lambda}\mathbf{V}^{-1})\, \mathbf{B}^{\rm T}
  = (\mathbf{BV})\,\bm{\lambda}\,(\mathbf{V}^{\rm T}\mathbf{B}^{\rm T})\nonumber\\
  &= (\mathbf{BV})\,\bm{\lambda}\,(\mathbf{BV})^{\rm T}.
\end{align}
If $(\mathbf{BV})^{\rm T} = (\mathbf{BV})^{-1}$, the expression above would be the eigenvalue decomposition of $\mathbf{C}_{\rm B}$, making it appear that the eigenvalues of $\mathbf{C}_{\rm B}$ are the same as those of $\mathbf{C}$, and the eigenvectors are simply those of $\mathbf{C}$, binned using $\mathbf{B}$.~At face value, this is sensible: if the eigenvectors were smooth on scales smaller than the `bin width' then they will be unchanged by -- essentially invariant to -- the binning.  

However, notice that
\begin{equation}
 \mathbf{BV} \, (\mathbf{BV})^{\rm T} = \mathbf{B}\, (\mathbf{VV}^{\rm T}) \, \mathbf{B}^{\rm T} = \mathbf{B}\,\mathbf{B}^{\rm T} 
\end{equation}
is not diagonal (i.e., although $\mathbf{V}$ is an orthonormal basis, $\mathbf{V}^{\rm T}=\mathbf{V}^{-1}$, the same is not true for $\mathbf{BV}$).~This means that we should {\em not} think of $\mathbf{BV}$ as being the eigenvectors.~If we use $\mathbf{V}_{\rm B}$ to denote the eigenvectors of $\mathbf{C}_{\rm B}$,  
\begin{equation}
    \mathbf{C}_{\rm B} = \mathbf{V}_{\rm B} \bm{\lambda}_{\rm B} \mathbf{V}_{\rm B}^{-1},
\end{equation}
then it is natural to ask:  How different are the vectors which make up $\mathbf{V}_{\rm B}$ from those of $\mathbf{BV}$? 

\subsection{Numerical analysis}

Heuristically, we expect that if the binning remains smaller than the typical size of features in the eigenvectors, then they will be unchanged by binning.~This should be particularly true for the primary `cosmic variance' dominated eigenvectors; the shot-noise dominated eigenvectors oscillate more, but we argued that they are not interesting anyway.~Therefore, we expect the estimates of the BAO distance scale and their uncertainties should not depend on how the TPCF was binned, provided this binning is not wider than the BAO feature itself.~(If the bins are too wide, they will not provide a good description of the BAO feature anyway.) 

\begin{figure}
\includegraphics[width=0.49\textwidth]{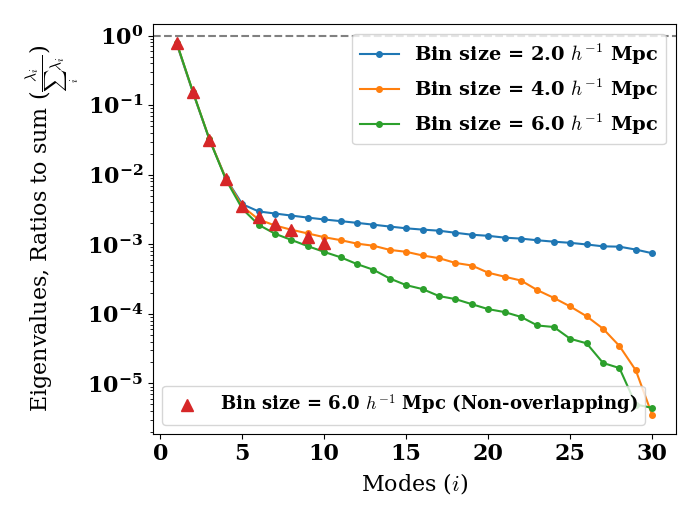}
\caption{\label{fig:eigvals_by_mode}Dependence of eigenvalues on bin-size and spacing.  Wider bins have smaller total variance, so we have normalized each eigenvalue to show its fractional contribution to the total. 
The first 4 normalized eigenvalues are very similar between all 3 curves, while the first 4 values for non-overlapping $6~h^{-1}$Mpc bins are essentially the same as the overlapping $6~h^{-1}$Mpc bins.}
\end{figure}

\begin{figure*}
\includegraphics[width=0.98\textwidth]{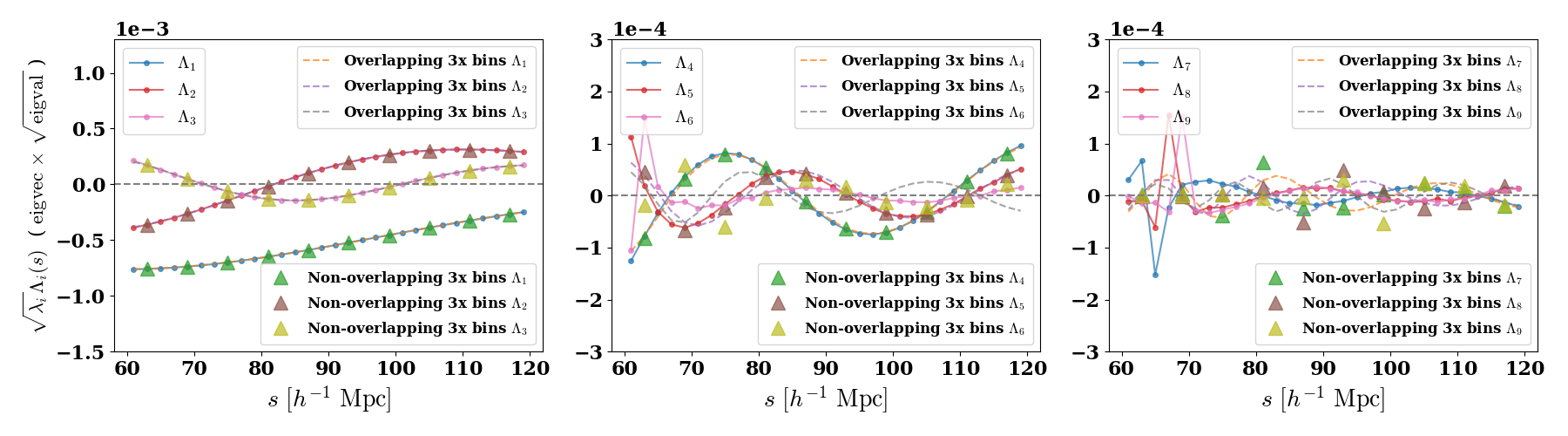}
\caption{\label{fig:eigvals_eigvecs_bins}First 9 eigenvectors $\sqrt{\lambda_i} \Lambda_i(s)$ at $z = 0.5057$, for bin-sizes $2~h^{-1}$Mpc and $6~h^{-1}$Mpc with the same bin centers. Larger symbols show the eigenvectors for the $6~h^{-1}$Mpc bins that do not overlap (so there are $3\times$ fewer bin centers, different from the overlapping bins). }
\end{figure*}

Figs.~\ref{fig:eigvals_by_mode} and~\ref{fig:eigvals_eigvecs_bins} show the result of two explicit tests.~The first increases the bin-size, but keeps the bin centers, and hence the number of bins, the same.~(As a result, neighboring bins are more correlated, but this just means that $\mathbf{C}_{\rm B}$, which has the same dimension as in the previous section, is less diagonal than the original $\mathbf{C}$.)~This corresponds to the $\epsilon\ne 1$ case mentioned previously (Section~\ref{sec:covariance_eigendecomposition}). 
Fig.~\ref{fig:eigvals_by_mode} shows the eigenvalues when we increase $\Delta r$ by factors of 2 and 3 ($\epsilon = 1/2$ and $\epsilon = 1/3$), respectively. The first few eigenvalues, which dominate the total variance, are indistinguishable from the original ones, but the more shot-noise dominated modes are affected.~In particular, for wider bins the shot-noise is smaller, so these shot-noise dominated modes contribute less to the total variance.~To remove the fact that the total variance is reduced, we normalized each set of eigenvalues by their total:~this shows explicitly that, as the bin-size increases, the shot-noise dominated modes contribute a smaller fraction of the total variance.  

The larger symbols show the eigenvalues when $\Delta r\to 3\Delta r$ but the bins do not overlap ($\epsilon = 1$).~In this case, there are $3\times$ fewer eigenvalues, so the total variance is obviously different.~Nevertheless, the fractional contribution of the first 10 modes to the variance is similar to that for the overlapping bins (of the same width).~Clearly, for the cosmic-variance dominated modes which contribute most to the total variance, the binning does not matter.  

Fig.~\ref{fig:eigvals_eigvecs_bins} shows that this is also true for the eigenvectors.~The big symbols show the case in which $\Delta r\to 3\Delta r$ but now the bins do not overlap (so $\epsilon=1$ even for this larger bin-size; for clarity, we do not show the intermediate case where $\Delta r\to 2\Delta r$).~Again, the eigenvectors which dominate the total variance (the first $\sim 5$) are indistinguishable from the original ones.~This is slightly non-trivial since now $C_{\rm B}$ is $10\times 10$ rather than $30\times 30$, but the leading eigenvectors are unchanged.~Hence, the uncertainties on distance scale estimates provided in the previous section will be unchanged:~they do {\em not} depend on the binning, at least for fiducial values of the shot-noise.  

More generally, if the convolution kernel which defines the binning does not erase features in the original (cosmic variance dominated) eigenvectors, then these eigenvectors will not depend on the exact bin shape.~E.g., this will certainly be true if the off-diagonal entries in Eq.~\ref{eq:matrixB} are less than unity.~Similarly, counting pairs in, e.g., Gaussian-like bins rather than in rectangles will not change our conclusions.  
  

\section{Discussion and Conclusions}\label{sec:conclusion}

We presented an eigen-decomposition of the Gauss-Poisson approximation to the covariance matrix of the two-point correlation function (Eq.~\ref{eq:TPCF_cov_mat_xi0}) and assessed the importance of the power spectrum-dominated modes that trace cosmic variance as opposed to the modes which are dominated by shot-noise.~For a fiducial cosmology and noise-levels that are consistent with current and next-generation surveys, the cosmic variance eigen-modes account for most of the total variance of the TPCF (Fig.~\ref{fig:eigvals_by_shot}).~They are also smoother than the shot-noise dominated modes (Fig.~\ref{fig:eigvals_eigvecs_vs_no_shot}), so they are insensitive to the modest changes in binning that are allowed if one wishes to resolve the BAO feature in the TPCF (Figs.~\ref{fig:eigvals_by_mode} and~\ref{fig:eigvals_eigvecs_bins}).

We argued that, as a result, the cosmic variance eigen-modes alone should provide a good description of the correlated residuals which result from fitting smooth functional forms to the measured TPCF.~We provided a simple (Eq.~\ref{eq:Delta_LP}) but accurate (Fig.~\ref{fig:sigma_BAO_estimates}) approximation for the uncertainty on the Linear Point estimate of the BAO distance scale which explains why the uncertainty is greater in the evolved field than in linear theory (Fig.~\ref{fig:sigma_BAO_estimates_linear}); allows one to quantify the gains from working with the reconstructed signal; and does not depend on the functional form fitted to the TPCF or the binning used.~It also provides insight into why the LP is more robust than the inflection point, and why both are more precise distance indicators than the peak or dip scales.~Perhaps most importantly, our approximation allows one to quantify the precision of the BAO distance scale estimate without having to generate a large ensemble of mock catalogs.~Therefore, it should be useful for estimating the gains in precision which come from making measurements in the reconstructed field (which are often quoted), after marginalizing over the unknown cosmological model (a step which is often ignored). 

Our approach also provides a realistic estimate of the precision of the zero-crossing scale (Fig.~\ref{fig:sigma_r0}), showing that the uncertainty on it depends on how steeply $\xi$ crosses zero.~Although it is considerably less precise than either $r_{\rm LP}$ or $r_{\rm infl}$ for realistic galaxy surveys, our analysis vastly simplifies the process of quantifying the synergy between joint analyses of $r_{\rm LP}$ and $r_0$ in the same dataset.

In practice, our analysis exploits the appearance of extended plateaus in plots of how the precision in the distance scale indicator varies as more (sorted) eigen-modes are included (c.f.~Figs.~\ref{fig:sigma_BAO_estimates},~\ref{fig:sigma_BAO_estimates_linear} and~\ref{fig:sigma_r0}), since it is these plateau heights that correspond to the uncertainty estimates that are returned from fitting basis functions to the measurements.~(If a plateau is not seen, then this is an indication that the contributions from the cosmic variance and shot-noise terms to the full uncertainty are not so easily separated, and this makes it harder to pinpoint the correspondence between our analysis and the standard one.)  These plateaus suggest that it should be possible to write down a prescription for determining the optimal number of eigen-modes which should be used in cosmological analyses. 

While we do not perform cosmological analysis to infer cosmological parameters or evaluate cosmological models in this work, our approach can be used to rapidly measure the BAO scale and its uncertainties accurately given a dataset.~This will enhance the utility of the LP that previous works have already shown: (1) the distance scale obtained from the LP is consistent with the standard template-fitting method, while being more model agnostic \cite{anselmi2018CF_standard_ruler} and (2) the resulting cosmological parameters inferred from the LP are consistent with the standard approach with 20 to 30\% larger uncertainties under a flat $\Lambda$CDM cosmology \cite{he2023_BOSS_LP}.~Furthermore, the rapid estimation of the BAO scale and its uncertainties means that performing multiple BAO analyses assuming a variety of different cosmological models will be possible at reasonable computational cost, which will be especially important in the light of the time-evolving dark energy results from \cite{DES-SN5YR_2024,DESI2024VI_Cosmology}. In future work, we hope to combine our eigen-mode analysis with the Bayesian framework of \cite{paranjape2022} utilizing the Laguerre reconstruction methodology of \cite{nikakhtar2021BAO_Laguerre,nikakhtar2021BAO_Laguerre_mock_catalogues}, for a completely model-agnostic BAO analysis.

\medskip

\begin{acknowledgments} 
JL was supported by DOE grant DE-FOA-0002424
and NSF grant AST-2108094.~FN gratefully acknowledges support from the Yale Center for Astronomy and
Astrophysics Prize Postdoctoral Fellowship.~AP and RKS are grateful to the ICTP, Trieste for its hospitality in summer 2024, and RKS is grateful to the EAIFR, Kigali for its hospitality when this work was completed.~The research of AP is supported by the Associates Scheme of ICTP, Trieste.
\end{acknowledgments}

\bibliography{apssamp}

\appendix
\section{Application to the zero-crossing scale}\label{sec:r0}

With some care, the zero-crossing scale, $r_0$, can also be used as a standard ruler \cite{Prada2011, bkBAO}.~In practice, estimating $r_0$ is similar to the Linear Point scale:~one fits the pair counts with suitably chosen basis functions before estimating the zero-crossing from the fit.  Therefore, the analysis in the main text can be modified to provide estimates of the uncertainty on $r_0$ as follows. 

If $s_0$ is the scale on which $\Xi(s) = 0$, and $r_0$ is where $\xi(r)=0$, we have
\begin{equation}
    0 = \xi(r_0) + \Delta_0\, d\xi/dr + \sum_i g_i [\Lambda_i(r_0) + \Delta_0\, d\Lambda/dr]
\end{equation}
where $\Delta_0 = s_0-r_0$.  
Since $\xi(r_0)=0$, dividing throughout by $\xi'(r_0)$ and then assuming  $\Lambda'(r_0)/\xi'(r_0) \ll 1$, yields 
\begin{equation}
  \frac{\Delta_0}{r_0} \approx 
   -\sum_i g_i\, \frac{\Lambda_i(r_0)}{r_0 \xi'(r_0)},
\end{equation}
so
\begin{equation}
  \left\langle\frac{\Delta_0^2}{r_0^2}\right\rangle 
  \approx \sum_i \lambda_i \,\left[\frac{\Lambda_i(r_0)}{r_0 \xi'(r_0)}\right]^2 .
\end{equation}
Notice that the uncertainty is smaller if $\xi'$ is bigger. This is sensible:~the zero-crossing of a curve is easier to detect if the curve is steep.~In practice, this means that more biased tracers are to be preferred for this measurement.  

\begin{figure}
\includegraphics[width=0.49\textwidth]{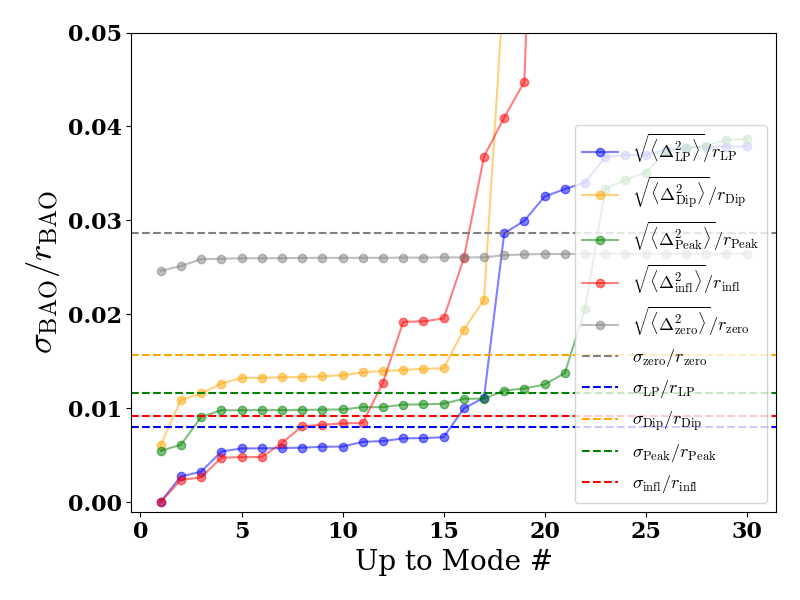}
\caption{\label{fig:sigma_r0}Same as Figure~\ref{fig:sigma_BAO_estimates}, but now showing the fractional uncertainties on the zero-crossing scale (grey) as well. }
\end{figure}

Figure~\ref{fig:sigma_r0} shows how well this compares to the standard approach (of fitting a polynomial to the pair counts over the range $100-160~h^{-1}$Mpc in bins of width $\Delta r=2~h^{-1}$Mpc) as the number of modes that is included in the sum increases.~The figure is in the same format as Figure~\ref{fig:sigma_BAO_estimates} in the main text, to highlight the similarity of the analysis.~The height of the plateau region defined by the grey symbols is our eigen-mode estimate of the uncertainty on $r_0$.~It tracks that from the standard (polynomial fitting) method quite well (grey dashed), showing that it provides reliable estimates for analyses of the zero-crossing scale.  

For $r_0$, there are two striking differences compared to our analyses of the scales ($r_{\rm LP}$ etc.) that we highlighted in the main text.~First, the plateau region -- which the main text argued is a proxy for the uncertainty on the distance scale -- is reached after only the first couple of eigen-modes.~For $r_{\rm LP}$ etc., more eigen-modes are need before convergence~to a plateau is seen, and then the plateau lasts for smaller range of eigen-modes before the variance increases again (mainly due to shot-noise dominated modes).~Second, the uncertainty on $r_0$ is much larger than on $r_{\rm LP}$ etc. This quantifies results which are obvious by eye in Ref.~\cite{bkBAO}.  In this respect, $r_{\rm LP}$ is the better distance scale indicator than $r_0$.~Nevertheless, there are synergies associated with having two distance scale estimates from the same measurement which we will explore elsewhere.

\end{document}